# Bloch Analysis of Finite Periodic Microring Chains

**Matteo Cherchi**

Pirelli Labs – Optical Innovation, viale Sarca 222, 20126 Milan, Italy

current address:

Department of Electrical Engineering, University of Palermo, 90128 Palermo, Italy.

E-mail: cherchi@cantab.net, FAX:

Abstract: We apply Bloch analysis to the study of finite periodic cascading of microring resonators. Diagonalization of the standard transfer matrix approach not only allows to find an exact analytic expression for transmission and reflection, but also to derive a closed form solution for the field in every point of the structure. To give more physical insight we analyze the main features of the transmission resonances in a finite chain and we give some hints for their experimental verification.

# 1. Introduction

In the last few years there has been a growing interest in optical periodic structures and to their analogies with their electronic counterpart. Periodic index modulation is the basic idea of Photonic Crystals [1], whereas periodic coupling to microresonators defines the so called side-coupled integrated spaced sequence of resonators (SCISSORs) [2, 3]. On the other hand periodic coupling of microresonators is the concept underlying Coupled Resonators Optical Waveguides [4, 5] (CROWs). They can be classified in to two main classes: traveling wave, like those featuring microrings as resonators, and standing wave, like those made of Fabry-Perot cavities in multilayer systems or defect cavities in photonic crystals. In this paper we will focus on a particular implementation of traveling wave CROWs, the microring chains (Fig. 1.a), where the resonators are coupled to each other.

Even though these kind of structure has been introduced mainly for nonlinear applications [5-7], we will focus on the solution of the linear problem of transmission, reflection and field distribution in a finite periodic chain.

In previous works [8, 9] the study of finite structures has been performed numerically, with a transfer matrix approach or analytically through a coupling of mode approach [10]. Also the envelope function approach [11] has been proposed, which makes use both of coupling of mode theory and Bloch analysis applied to periodic multilayers, in the limit of low index contrast. Only semi-infinite structures have been treated analytically through a Bloch mode analysis [8], and the tight-binding approximation [4, 9, 12], that treats each resonator as only slightly perturbed by its neighbour, so that the modes of the whole structure can be written in terms of single resonator modes. In particular this last approach not only is valid for small coupling only, but also can not be straightforwardly extended to study light propagation in finite periodic media.

In the last few year, the effort to get an analytic solution for the linear propagation in a finite periodic multilayer structure, has produced two somewhat complementary results, that we will introduce using an analogy. Let us consider a Fabry-Pérot: it can be seen both as an open cavity, i.e. a cavity with its proper "standing waves" (the quasinormal modes [13]) or, on the other hand, as a cascading of two mirrors that reflect progressive and regressive traveling waves, i.e. the eigenmodes of the medium that they surround. These two points of view can be applied also to periodic structures: the first one by finding the proper quasinormal modes [14], the second one by replacing the plane waves with the Bloch waves, that is what we have done in a recent paper [15].

In perfect analogy with this last work on periodic multilayers, we will show that Bloch analysis is a very powerful tool also for the study of finite periodic microring chains. It allows to derive exact analytic expressions for transmission, reflection and for the field inside the structure.

## 2. Transfer Matrix

The system under study is a finite perfectly periodic chain of identical microrings, as shown in Fig. [1]. The ring length is $2\Lambda$ and the waveguide propagation constant $\beta$ is assumed to be constant all over the ring section. The rings are all equally spaced, so that coupling between adjacent rings is always the same.

The system is topologically equivalent to the ideal Fabry-Perot structure in Fig. 1 with mirrors of null thickness and cavity length $\Lambda$. This topological equivalence allows to reduce to a purely 1-D problem, identifying two symmetric points of the ring (one in the upper and one in the lower path) with a single point of a straight line. In this way we can always expand the field $E_n(z)$ of the $n$th ring on the local basis of forward and backward propagating plane waves $\{E_n^f, E_n^b\}$, centered in the middle of the period, that is $E_n^f(z) \equiv \exp[-i\beta(z-n\Lambda)]$ and $E_n^b(z) \equiv \exp[i\beta(z-n\Lambda)]$. Expanding the field in this basis, we can write

$$E_n(z) = a_n E_n^f(z) + b_n E_n^b(z), \qquad (1)$$

where the complex numbers $a_n$ and $b_n$ must be constant, because the plane waves are the eigenstates of uniform media. Since $a_n$ and $b_n$ completely determine the state, we can represent the field in the $n$th ring through the complex vector

$$\mathbf{\Phi}_n \equiv \begin{pmatrix} a_n \\ b_n \end{pmatrix}. \qquad (2)$$

Symmetry of the couplers requires the transmittance $it$ between adjacent cavities (i.e. the coupling) to be purely imaginary, and the reflectance $r$ to be real [16]. Assuming negligible losses, the matrix propagating the field from the middle points of the $n$th ring to the middle points of the $(n+1)$th ring will be

$$\mathbf{P}_\Phi = \frac{1}{it}\begin{pmatrix} -e^{-i\beta\Lambda} & r \\ -r & e^{i\beta\Lambda} \end{pmatrix}, \qquad (3)$$

so that $\mathbf{\Phi}_{n+1} = \mathbf{P}_\Phi \mathbf{\Phi}_n$.

## 3. Bloch Modes

The most efficient way to study a physical system is representing its state on the basis of its eigenmodes. They allow to determine the state in the whole structure by simply knowing it in a single point. The eigenmodes of a periodic system are nothing but the Bloch modes. To find their representation in our formalism we have just to diagonalize $\mathbf{P}_\Phi$. Its eigenvalues are easily found to be $\lambda_\pm = \exp(\pm ik\Lambda)$, where

$$\cos(k\Lambda) \equiv \frac{\sin(\beta\Lambda)}{t}, \tag{4}$$

that defines $k$ as the Bloch propagation constant. They correspond to the eigenvectors

$$\mathbf{\Phi}_\pm \equiv \begin{pmatrix} a_\pm \\ b_\pm \end{pmatrix} \tag{5}$$

which elements must satisfy the conditions $ua_+ = rb_+$ and $ra_- = ub_-$, where we have defined $u \equiv \cos(\beta\Lambda) - \sqrt{\cos^2(\beta\Lambda) - r^2}$. Clearly they obey the equation $\mathbf{P}_\Phi \mathbf{\Phi}_\pm = \exp(\pm ik\Lambda) \mathbf{\Phi}_\pm$.

This means that, up to an overall phase, the Bloch modes can be written as

$$E^\pm(z) \equiv \exp[\pm in(z)k\Lambda][a_\pm E_n^f(z) + b_\pm E_n^b(z)] \equiv \exp[\pm in(z)k\Lambda]E_n^\pm(z) \tag{6}$$

where we have expressed the period number as a function of $z$, that is $n(z) = \text{int}(z/\Lambda + \tfrac{1}{2})$ and 'int' means integer part. They can also be rewritten in the standard Bloch form

$$E^\pm(z) = e^{\pm ikz}\{\exp[\mp ik(z - n(z)\Lambda)]E_n^\pm(z)\} \tag{7}$$

where the expression in parentheses is clearly $\Lambda$–periodic.

Notice that, from Eq. (4), the stop bands correspond to $\Delta \equiv t^2 - \sin^2(\beta\Lambda) = \cos^2(\beta\Lambda) - r^2 < 0$.

Since the Bloch modes are the eigenmodes of the periodic structures, they are the most convenient basis to represent the field as:

$$E(z) = x_n E_n^+(z) + y_n E_n^-(z) = x_0 E^+(z) + y_0 E^-(z), \qquad (8)$$

where $x_n$ and $y_n$ are complex coefficients. We like to point out that, thanks to the intrinsic nature of the Bloch modes, in Eq. (6) we have been able not only to find a local basis $\{E_n^+(z), E_n^-(z)\}$, but also a closed form for a global basis $\{E^+(z), E^-(z)\}$.

In the new local basis the vector representing the field will be

$$\Psi_n \equiv \begin{pmatrix} x_n \\ y_n \end{pmatrix} \equiv \mathbf{W}\Phi_n \qquad (9)$$

where

$$\mathbf{W}^{-1} \equiv \begin{pmatrix} a_+ & a_- \\ b_+ & b_- \end{pmatrix}, \quad \mathbf{W} = \begin{pmatrix} \bar{b}_- & -\bar{a}_- \\ -\bar{b}_+ & \bar{a}_+ \end{pmatrix},$$

and $\bar{a}_\pm \equiv a_\pm / \det(\mathbf{W}^{-1})$, $\bar{b}_\pm \equiv b_\pm / \det(\mathbf{W}^{-1})$. Clearly, in this representation, the matrix linking two adjacent periods will be $\mathbf{P}_\Psi \equiv \mathbf{W}\mathbf{P}_\Phi\mathbf{W}^{-1} = \text{diag}[\exp(ik\Lambda), \exp(-ik\Lambda)]$.

Notice that whenever $\Delta = 0$ we have $\det(\mathbf{W}^{-1}) = 0$. This corresponds to the band edges, where $u^2 = \cos^2(\beta\Lambda) = r^2$ and $\cos(k\Lambda) = 1$. The two eigenvectors are degenerate and $\mathbf{W}^{-1}$ is not invertible. In this case the Bloch modes cease to be a basis for the system.

## 4. Boundary Conditions

Equation (8) means that we can express the field in every point of the structure by simply knowing the Bloch coefficients $x_0$ and $y_0$ at the beginning. To determine $\mathbf{\Psi}_0$ we have simply to impose the boundary conditions. In terms of the input and output states $\mathbf{\Phi}_0$ and $\mathbf{\Phi}_N$ they read

$$\begin{cases} \mathbf{\Psi}_0 = \mathbf{W}\mathbf{\Phi}_0 \\ (\mathbf{P}_\Psi)^N \mathbf{\Psi}_0 = \mathbf{W}\mathbf{\Phi}_N \end{cases}. \tag{10}$$

If we now assume, as usual, to couple light in only one port of the 0th period, we get

$$\mathbf{\Phi}_0 \equiv \begin{pmatrix} 1 \\ \rho \end{pmatrix} \text{ and } \mathbf{\Phi}_N \equiv \begin{pmatrix} \tau \\ 0 \end{pmatrix}$$

where $\rho$ and $\tau$ are the reflectance and transmittance of the whole structure, i.e. two complex numbers to be determined together with $x_0$ and $y_0$. The linear system (10) can be easily solved to give:

$$\begin{cases} \tau = \Omega_N (a_+ b_- - a_- b_+) \\ \rho = -2i\Omega_N \sin(Nk\Lambda) b_+ b_- \\ x_0 = e^{-iNk\Lambda} \Omega_N b_- \\ y_0 = -e^{iNk\Lambda} \Omega_N b_+ \end{cases}, \tag{11}$$

where $\Omega_N \equiv [e^{-iNk\Lambda} a_+ b_- - e^{iNk\Lambda} a_- b_+]^{-1}$.

This is a very general result, which holds for any 1-D bi-directional periodic system. To specialize to the case under study, we can choose $a_+ = b_- = r$ and $a_- = b_+ = u$, to give

$$\begin{cases} \tau = \Omega_N(r^2 - u^2) \\ \rho = -2i\Omega_N \sin(Nk\Lambda)\,u\,r \\ x_0 = e^{-iNk\Lambda}\Omega_N\,r \\ y_0 = -e^{iNk\Lambda}\Omega_N\,u \end{cases} \qquad (12)$$

and $\Omega_N = [e^{-iNk\Lambda}r^2 - e^{iNk\Lambda}u^2]^{-1}$.

This is an exact closed form solution for the problem of a finite periodic chain. Notice that at the band edges $\tau \to t/[t - iN\cos(\beta\Lambda)]$ and $\rho \to -irN/[t - iN\cos(\beta\Lambda)]$ are well behaved functions, whereas $x_0$ and $y_0$ diverge, due to the degeneracy of the basis.

From Eq. (12) and Eq. (9) we can also recover a closed form for the plane wave coefficients $\mathbf{\Phi}_n = \mathbf{W}^{-1}(\mathbf{P}_\Psi)^n \mathbf{\Psi}_0$ or

$$\begin{cases} a_n = \Omega_N/\Omega_{N-n} \\ b_n = -2i\Omega_N \sin[(N-n)k\Lambda]\,u\,r \end{cases} \qquad (13)$$

Even though we have worked in a basis that is degenerate at the band edges, when going back to the plane waves basis, we find that both $a_n \to [t - i(N-n)\cos(\beta\Lambda)]/[t - iN\cos(\beta\Lambda)]$ and $b_n \to -ir(N-n)/[t - iN\cos(\beta\Lambda)]$ are well behaved functions. Equation (13), together with Eq. (1), represents an exact analytic expression for the field in all forward and backward propagating branches of the rings. To our knowledge this is the first time that exact analytic expressions are calculated for the field in the periodic chain. In Ref. [8] only transmission and reflection of a finite periodic structure have been calculated analytically, and without making use of Bloch analysis, which is applied to infinite structures only. We have shown that Bloch analysis is superior also for finite structures, since it allows to calculate a closed form for the field in every point of the structure.

## 5. Numerical Example

As an example we have considered a chain of 11 rings (or $N=12$ periods) with constant reflectivity $r=\sqrt{0.2}$ and constant effective index $n_{eff}=1.8$ for all wavelength. We have chosen the ring length to be 10 optical cycles long at $\lambda_0=1.55\mu m$. We like to point out that these strongly coupled resonators are far away from the tight binding approximation. Amplitude and phase of the transmittance $\tau$ in Eq. (12) are shown in Fig. 2, together with the modulus of the Bloch coefficients $x_0$ and $y_0$. Notice the 11 peaks in the transmission window, corresponding to the beating of the Bloch wave number $k$ with a multiple of the reciprocal lattice wave number $K=2\pi/\Lambda$. They are the resonances of the Bloch modes in the finite structure (see Fig. 3). Resonance, i.e. total transmission, is the only case so that the input and output state are the same (up to an overall phase). This requires that, after $N$ periods, the phase difference between the two Bloch modes must return to its initial value. In terms of (real) Bloch wave numbers this means $Nk\Lambda = p\pi$ with $p \in \mathbb{N}$. On the other hand, from Eq. (4) it is also clear that $k\Lambda = m\pi$ ($m \in \mathbb{N}$) corresponds to a band edge (dashed lines in Fig. 2). So in a transmission window between the $m$th and the $(m+1)$th band edge, the peaks correspond to $k\Lambda = (m+q/N)\pi$ with $1 \le q \le N-1$. and beating length $L_b \equiv 2\pi/|k-mK| = N\Lambda/q$. Figures 3.a, 3.b, and 3.c show the amplitudes of the forward and backward plane waves corresponding to the resonances $q=1$, $q=2$, and $q=N/2$ of the central transmission band in Fig. 2. The field patterns repeat symmetrically for the transmission peaks on the left of $\lambda_0=1.55\mu m$. This is better understood when considering the beating with the wave vector associated with the closer band edge, namely $(m+1)K$. Notice the remarkable field pattern related to the central peak (present for odd number of rings only) featuring 5 rings with null backward field.

Regarding Fig. 2.c, notice not only the asymptotes at the band edges, but also that in the central transmission band $|x_0| > |y_0|$, whereas in the side transmission bands $|x_0| < |y_0|$. This can be understood when considering that $x_0$ is associated with the eigenvalue $\exp(ik\Lambda)$ and $y_0$ corresponds to $\exp(-ik\Lambda)$. If we look at the bands in the first Brillouin zone (Fig. 4), we clearly see that if below a given band gap a certain $\bar{k} > 0$ is associated with a progressive Bloch wave, below the same band gap it will correspond to a regressive wave. Since we are exciting the system from left to right, the progressive component must be always greater than the regressive one, and this explains why $x_0$ and $y_0$ must exchange their role in adjacent transmission bands. The same is true for the band gaps, where the mode decreasing with $z$ clearly dominates.

On the other hand, in the transmission band the amplitude of the regressive wave is not negligible. This means that the response of a finite structure cannot be approximated as the excitation of a progressive Bloch wave only (that is approximating it as a semi-infinite structure). This would be like neglecting the backward component in a Fabry-Perot.

We also point out that, in the band structure, $\omega = 0$ does not correspond to $k = 0$. This is because, even for $r = 0$, a ring chain features a periodic $\pi/2$-phase shift due to the imaginary transmittance of the couplers, which translates the dispersion relation of $\pm K/4$. This is clear in Eq. (4), where a cosine equals a sine.

## 6. Applications and Extensions

Clearly having an analytic expression for the transmission of a periodic chain, allows to easily calculate time delay, group velocity (that is the inverse of the density of modes) and chromatic dispersion. Furthermore this allows to easily solve the problem of a periodic structure with a defect, since the two periodic sections surrounding the defect can be treated as concentrated mirrors with transmittance and reflectance given by Eq. (12).

More remarkably, having an analytic expression for the field in every point of the structure allows to solve more complicated problems, like, for example, second harmonic generation in the undepleted pump regime, by simply imposing the proper boundary conditions. The procedure is identical to the one we have proposed for periodic multilayers [15].

The only physical difference between a plane wave structure and a guided traveling wave structure is that, in the last case, forward and backward waves follow physically distinct paths. This can be relevant in the study of certain non-linearities. In a sense this feature implies that, for a microring chain, Bloch modes are a very powerful mathematical tool rather than a "real" physical entity.

The general form of Eq. (11) allows also to easily extend our approach to more complicated periods, featuring many different rings and/or coupling coefficients. What is needed is just the calculation of the eigenvectors $\Phi_{\pm}$ of the transfer matrix linking adjacent periods.

Another straightforward extension could account for absorbing media [17]. In this case the Bloch propagation constant will be in general a complex number.

We also point out that our formalism is very specific for perfectly periodic structures, and, by construction, cannot effectively treat apodized structures [18, 19], neither straightforwardly take in account the effect of random errors [20] in the chain.

## 7. Practical implementations

A nice experimental verification of the bright and dark paths in the resonance patterns shown in Fig. 3 (i.e. a qualitative measure of the square modulus of the coefficients $a_n$ and $b_n$ in Eq. 13) could be easily done by collecting the light scattered from the top of a chain similar to that proposed in our numerical example (where 12 periods have been chosen since 12 is divisible by 2, 3, 4 and 6, but a simpler realization could be done with 6 periods). Clearly the launched radiation should come from a tunable laser with linewidth, tenability and stability able to resolve the resonance peaks of the structure. This can be easily achieved if the quality factor $Q$ of the resonances is not very high, a condition that ensures also that the effects of material absorption and ring radiation losses can be neglected. But low $Q$ means, in turn, small ring radii and/or strong coupling (that is high transmittance $t$). Now suppose to keep the rings as close as possible (i.e. at the limit of the spatial resolution of the fabrication technique). Since, in general, the smaller the ring radius the smaller is the coupling that can be achieved (smaller radii not only mean shorter interaction lengths, but since they are achieved with higher index contrasts, they also give less coupling per unit length) a trade off must be found, depending on the ratio between material loss and radiation loss.

In the last few years a lot of effort [21] has been done to fabricate low loss rings with small radii and high quality factor $Q$, which are intended mainly for filters in telecom applications. Many materials have been proposed and tested, including organic materials [22], semiconductors [23] and oxides [24].

A particularly promising material is the Silicon Oxinitride (SiON) [25, 26], whose refractive index can be engineered in the range from that of Silicon Oxide ($SiO_2$) to that of Silicon Nitride ($Si_3N_4$).

We believe that these materials, that have been proposed to minimize radiation losses thanks to high index contrast waveguides, are the best candidate to fabricate the proposed low $Q$ structure. Anyway, as explained above, care must be taken in choosing the waveguide index contrast, that must not be too high in order to have strongly coupled resonators.

With current technologies, fabrication of an optimized working structure shouldn't be very difficult, since low *Q* means also high tolerance to fabrication errors.

## 8. Conclusions

In departure from previous works, we have shown that Bloch analysis, which is equivalent to the diagonalization of the transfer matrix formalism, allows an analytical treatment of finite periodic microring chains. The results are clearly valid for any bidirectional 1-D problem.

Next we have analyzed the occurrence of transmission peaks and interpreted the correspondent field patterns in terms of Bloch mode resonances.

We also suggest extensions and applications of the proposed formalism.

Finally we propose an interesting linear experiment that should be easily implemented in the strong coupling regime, usually not considered because it is not relevant in the context of nonlinear optics.

**Figure Captions**

Fig. 1.a) A chain of *N*-1 identical rings and b) its topological 1-D equivalence. To each point of the z axis correspond two symmetric points of the rings: one in the forward branch and the other in the backward branch.

Fig. 2. a) Transmission and b) phase of a 11 ring chain. Transmission windows are staggered by band gaps, which edges are plotted as dashed line. Each window is characterized by 11 resonant peaks.

Fig. 3. . Forward and backward power corresponding to the peaks of Fig. 2: a) q=1, with beating length *N*Λ; b) q=2, with beating length *N*Λ/2; c) q=6, with beating length 2Λ.

Fig. 4 Bands in the first Brillouin zone, where the Bloch wave vector *k* is normalized to *K*/2 and the angular frequency $\omega$ is normalized to $\omega_0 \equiv 2\pi c/\lambda_0$.

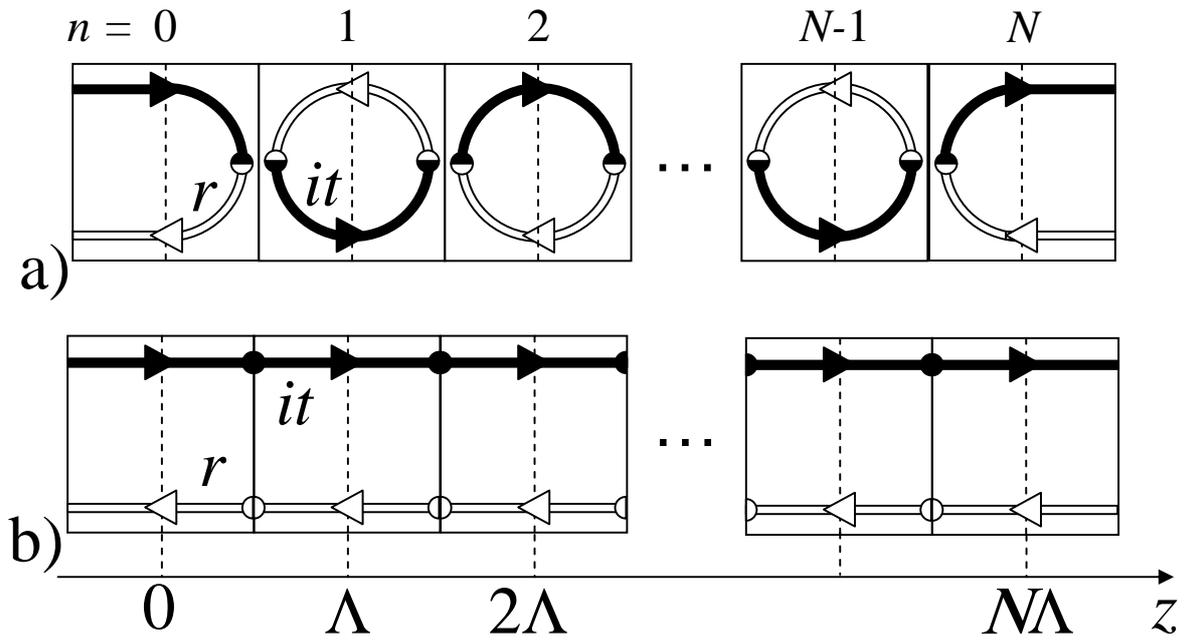

Fig. 1. a) A chain of *N*-1 identical rings and b) its topological 1-D equivalence. To each point of the z axis correspond two symmetric points of the rings: one in the forward branch and the other in the backward branch.

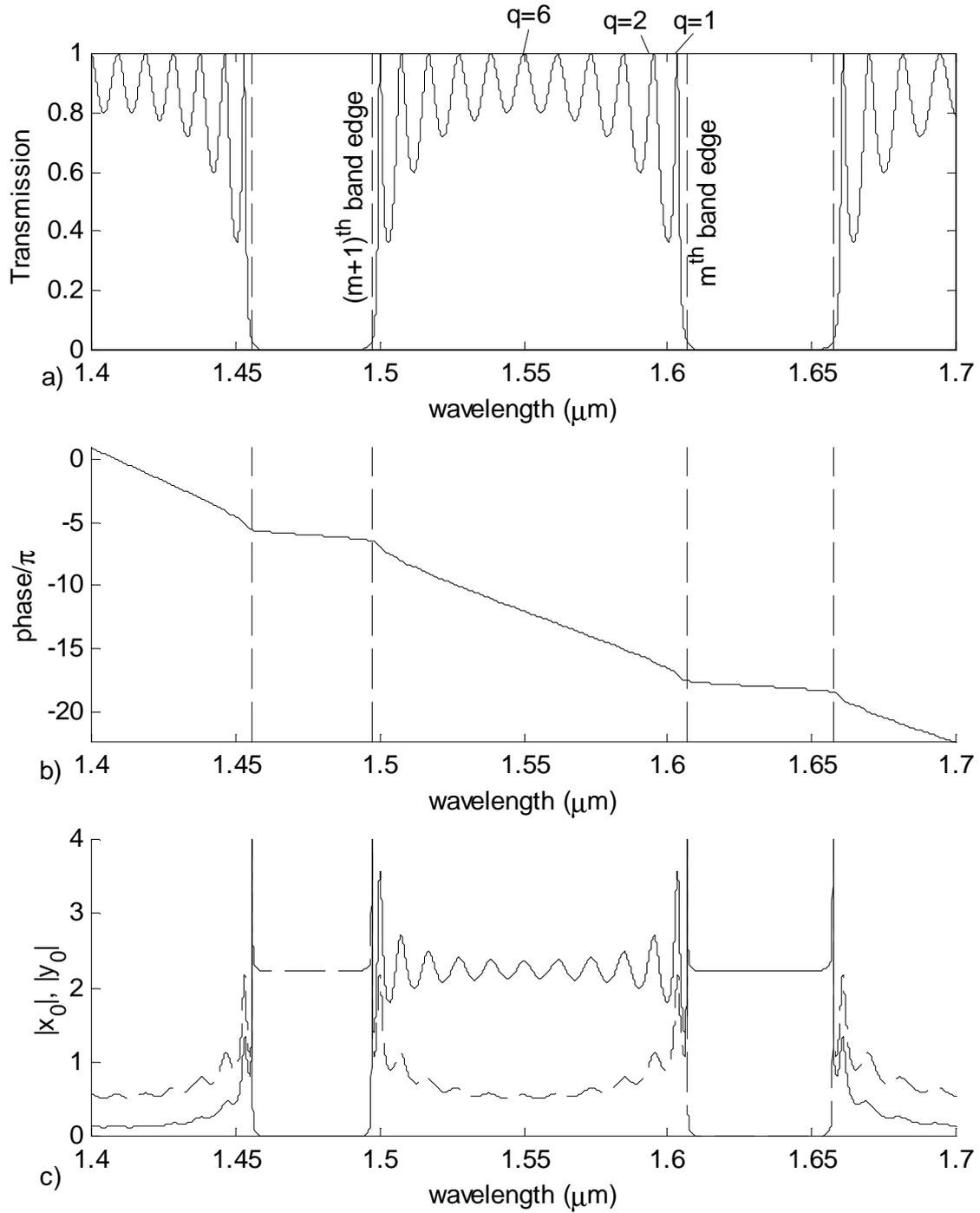

Fig. 2. a) Transmission and b) phase of a 11 ring chain. Transmission windows are staggered by band gaps, which edges are plotted as dashed line. Each window is characterized by 11 resonant peaks. Plot c) shows the moduli of the Bloch coefficients $x_0$ (solid line) and $y_0$ (dashed line).

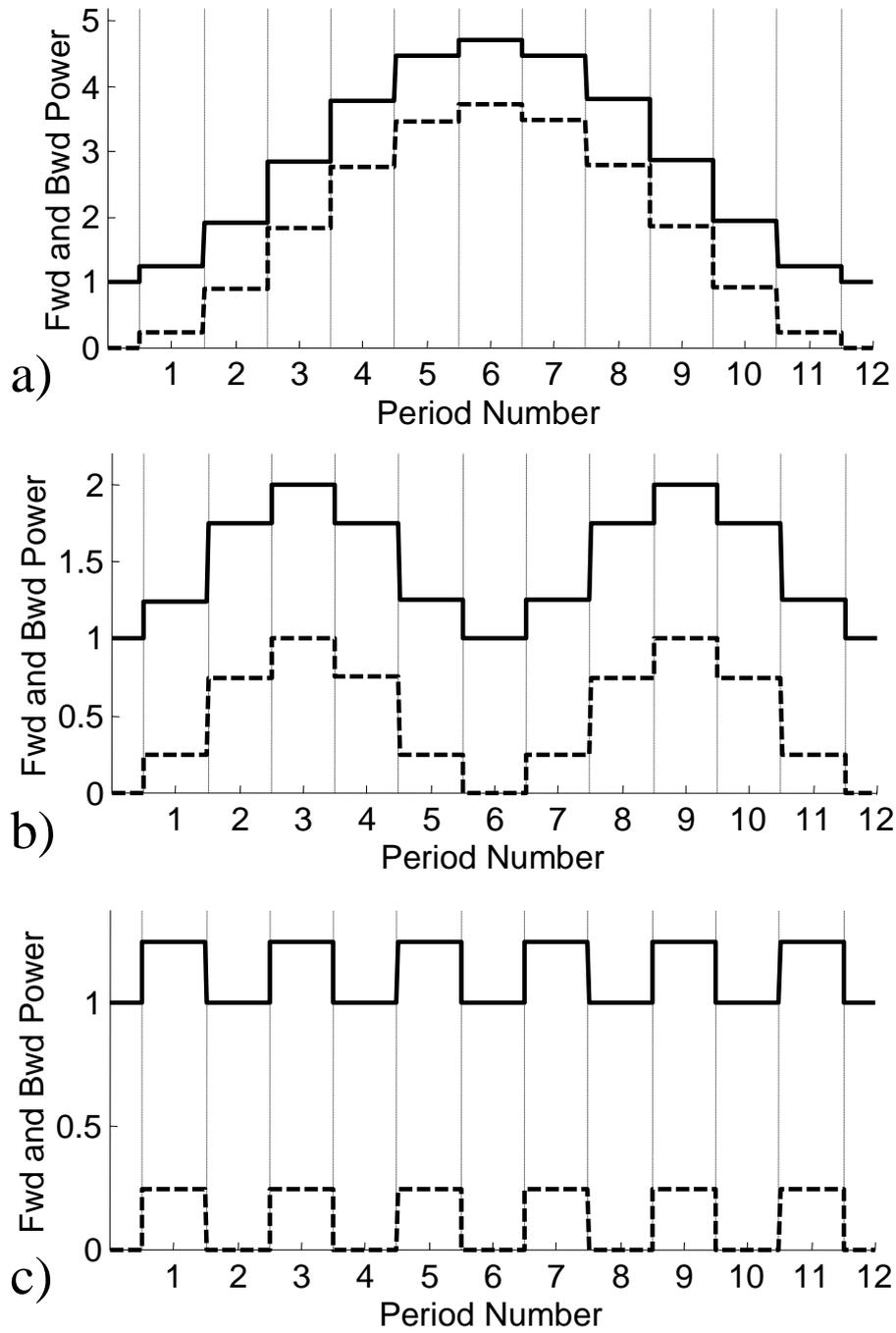

Fig. 3. Forward and backward power corresponding to the peaks of Fig. 2: a) q=1, with beating length $N\Lambda$; b) q=2, with beating length $N\Lambda/2$; c) q=6, with beating length $2\Lambda$.

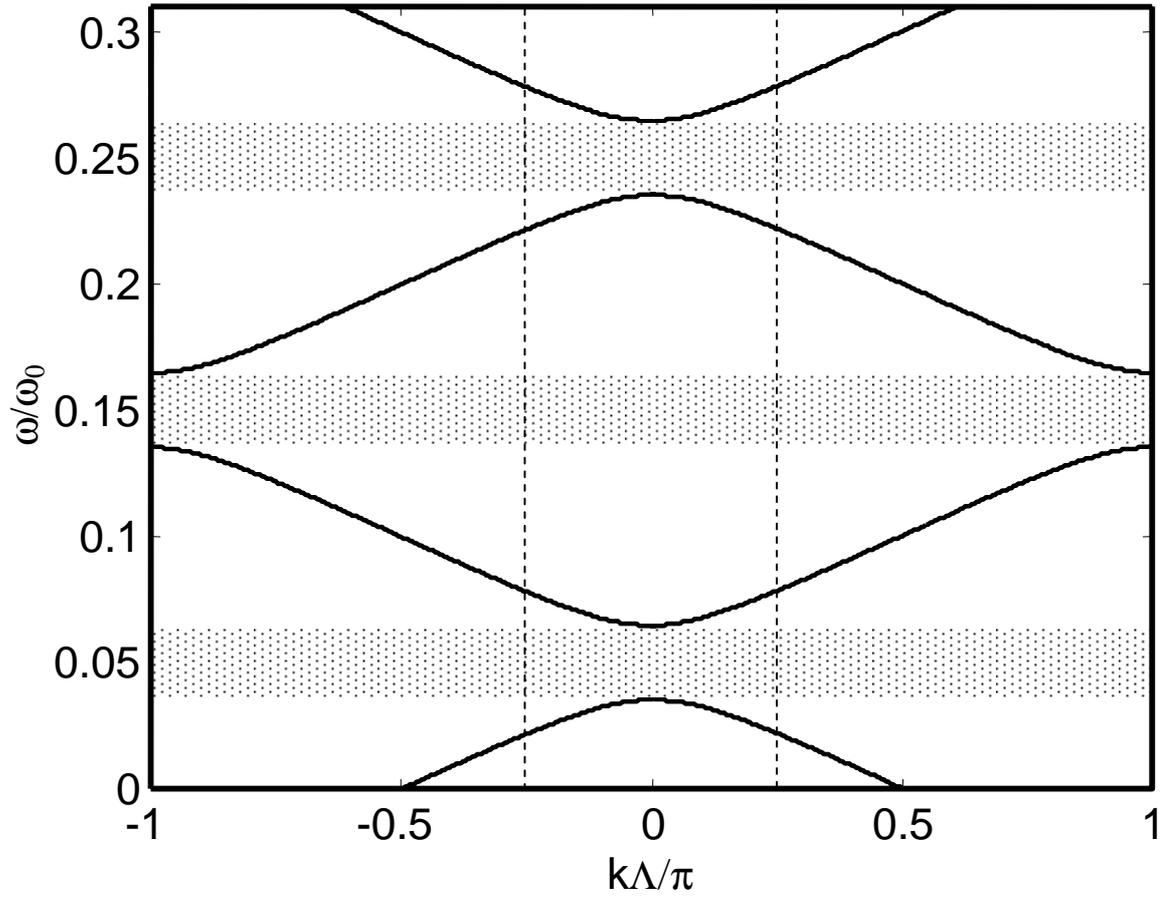

Fig. 4. Bands in the first Brillouin zone, where the Bloch wave vector $k$ is normalized to $K/2$ and the angular frequency $\omega$ is normalized to $\omega_0 \equiv 2\pi c/\lambda_0$.